\documentclass[aps,prl,twocolumn,superscriptaddress,nopacs,nofootinbib]{revtex4-1}

\usepackage{amsmath,amssymb,latexsym}
\usepackage{graphicx}
\usepackage{hyperref}
\usepackage{txfonts}
\usepackage{mathrsfs}
\usepackage{color}
\DeclareSymbolFont{symbols}{OMS}{cmsy}{m}{n}
\DeclareSymbolFont{largesymbols}{OMX}{cmex}{m}{n}
\renewcommand{\bm}[1]{\boldsymbol #1}

\begin{document}

\title{
Exact Out-of-Time-Ordered Correlation Functions for an Interacting Lattice Fermion Model
}

\author{Naoto Tsuji}
\affiliation{RIKEN Center for Emergent Matter Science (CEMS), Wako 351-0198, Japan}
\author{Philipp Werner}
\affiliation{Department of Physics, University of Fribourg, 1700 Fribourg, Switzerland}
\author{Masahito Ueda}
\affiliation{RIKEN Center for Emergent Matter Science (CEMS), Wako 351-0198, Japan}
\affiliation{Department of Physics, University of Tokyo, Hongo, Tokyo 113-0033, Japan}

\begin{abstract}
Exact solutions for local equilibrium and nonequilibrium out-of-time-ordered correlation (OTOC) functions are obtained
for a lattice fermion model with on-site interactions, namely the Falicov-Kimball (FK) model, 
in the large dimensional and thermodynamic limit. 
Our approach is based on the nonequilibrium dynamical mean-field theory
generalized to an extended Kadanoff-Baym contour. 
We find that the density-density OTOC is most enhanced 
at intermediate coupling around the metal-insulator phase transition.
In the high-temperature limit, the OTOC remains nontrivially finite and interaction-dependent, 
even though dynamical charge correlations probed by an ordinary response function are completely suppressed.
We propose an experiment to measure OTOCs
of fermionic lattice systems including the FK and Hubbard models
in ultracold atomic systems.
\end{abstract}


\date{\today}

\pacs{
}

\maketitle

There is a growing interest in the scrambling and spreading of information in quantum many-body systems
in wide areas of physics ranging from condensed matter to black holes
\cite{HaydenPreskill2007,SekinoSusskind2008,ShenkerStanford2014a,
ShenkerStanford2014b,Kitaev2014,Kitaev2015,ShenkerStanford2015,Hosur2016,MaldacenaShenkerStanford2016,Swingle2016,
Danshita2016,Zhu2016,Yao2016,Huang2016,Fan2016,Shen2016,Chen2016,SwingleChowdhury2016,He2016,Garttner2016,Li2016,Aleiner2016}.
A useful measure to diagnose the sensitivity of time-evolving quantities on the initial condition
is the out-of-time-ordered correlation (OTOC) function \cite{LarkinOvchinnikov1968} of two operators $W$ and $V$,
\begin{align}
C(t)
&=
-\langle [W(t),V(0)]^2\rangle.
\label{OTOC}
\end{align}
In the semiclassical picture, if we choose $W=p_j, V=p_k$,
$C(t)\sim\langle (\partial p_j(t)/\partial q_k(0))^2\rangle$ 
($p_j$ and $q_k$ are canonical momenta and coordinates), so that OTOCs reflect how
the system is scrambled and the initial-condition dependence is amplified.
In chaotic systems, OTOCs are expected to grow exponentially in time (butterfly effect).
It has been conjectured based on the holographic principle
that there is a universal bound for the exponential growth rate of OTOCs (Lyapnov exponent)
\cite{MaldacenaShenkerStanford2016}.
Recently, the OTOC has been analytically evaluated for the Sachdev-Ye-Kitaev (SYK) model,
a model of fermions having all-to-all four-body random interactions without hopping
\cite{SachdevYe1993,Kitaev2015,Sachdev2015,Jensen2016,PolchinskiRosenhaus2016,MaldacenaStanford2016}.
It was found to show exactly such a chaotic behavior with the bound saturated \cite{Kitaev2015}.

An immediate question is how OTOCs grow 
in the lattice fermion models with short-range interactions
that are typically used to describe strongly correlated condensed matter systems.
To evaluate Eq.~(\ref{OTOC}), one needs to compute quantities like $\langle W(t)V(0)W(t)V(0)\rangle$,
which are incompatible with the usual time-ordered sequence of operators
on the Keldysh or Kadanoff-Baym contour $\mathcal C$ ($0\to t\to 0\to -i\beta$, $\beta$ is the initial inverse temperature) \cite{Schwinger1961, Keldysh1964, KadanoffBaymBook}.
This is in contrast to ordinary response functions.
For example, a nonlinear optical susceptibility is given by a combination of current correlators such as
$\langle [[[j(t),j(t')],j(t'')],j(t''')]\rangle$ with a causality constraint $t\ge t'\ge t''\ge t'''$ \cite{ButcherCotter}, 
which can always be defined on $\mathcal C$.
Due to the unconventional ordering, OTOCs have not been much studied for correlated lattice fermion models.
Up to now, OTOC functions have been calculated mostly by exact diagonalization for spin
\cite{Swingle2016,Yao2016,Huang2016,Fan2016,He2016} and boson \cite{Shen2016} systems
of small size.

In this Letter, we generalize the nonequilibrium dynamical mean-field theory (DMFT)
\cite{FreericksTurkowskiZlatic2006,noneqDMFTreview}
to an extended Kadanoff-Baym contour,
which allows one to calculate OTOC functions for lattice fermion models in the infinite-dimensional and thermodynamic limit.
We apply this technique to the Falicov-Kimball (FK) model 
\cite{FalicovKimball1969,FreericksZlatic2003},
which admits an exact solution due to infinitely many conserved quantities. 
The FK model exhibits intriguing properties
such as a metal-insulator transition, non-Fermi liquid behavior \cite{FreericksZlatic2003}, 
and Anderson localization in two dimensions \cite{Antipov2016}.
Its nonequilibrium aspects have also been studied
\cite{FreericksTurkowskiZlatic2006,Freericks2008,EcksteinKollar2008,TsujiOkaAoki2008,TsujiOkaAoki2009,Canovi2014,Herrmann2016}.
We show that OTOCs provide yet another insight into physics of this model,
that cannot be obtained from ordinary response functions.
Finally, we propose an experimental scheme with ultracold atoms
to measure OTOCs in a fermion model.

\begin{figure}[t]
\begin{center}
\includegraphics[width=6cm]{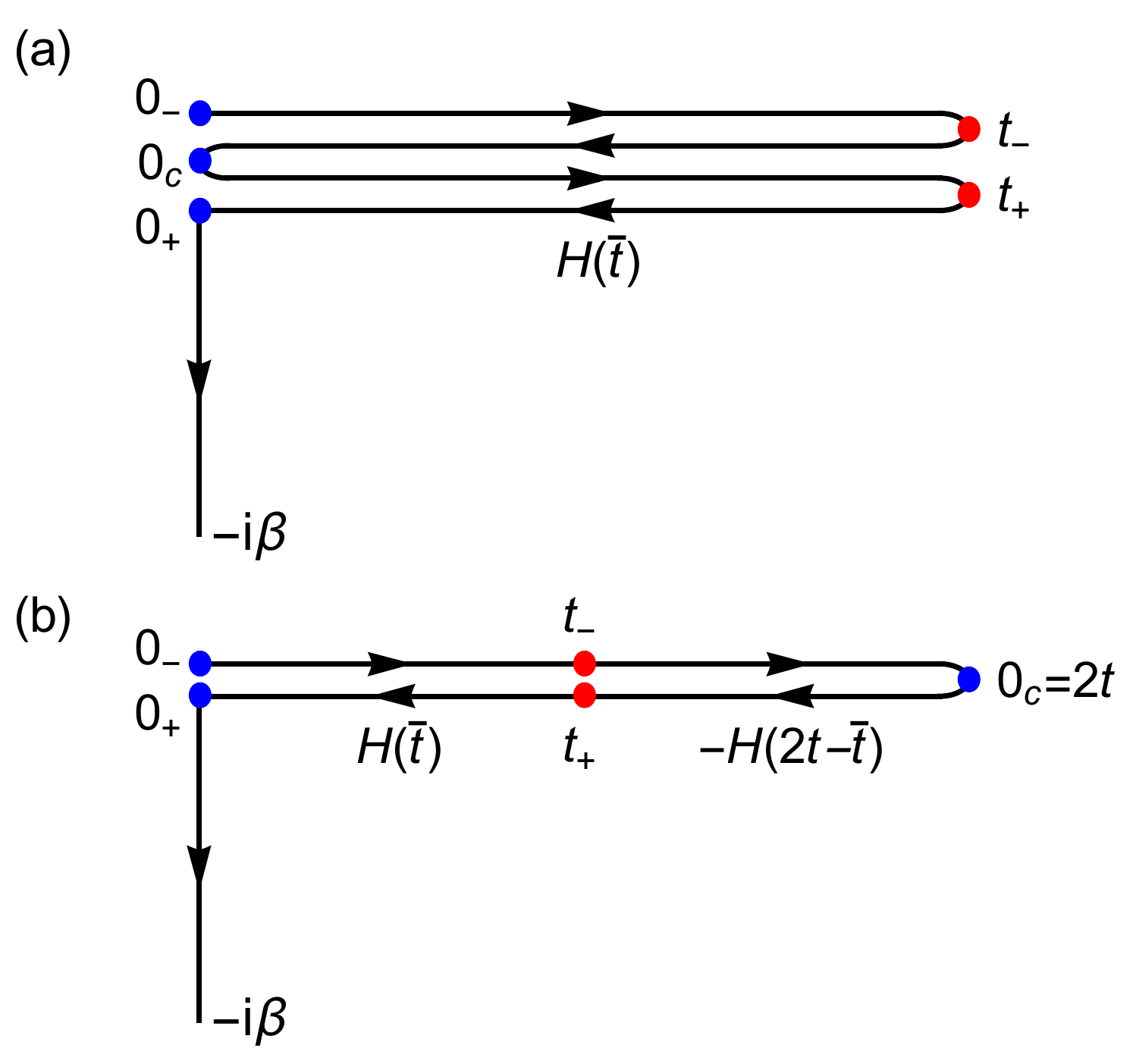}
\caption{Two types of extended [doubly folded (a) and singly folded (b)] Kadanoff-Baym contours $\tilde{\mathcal C}$,
which are equivalent. In (a), the system evolves with the Hamiltonian $H(\bar t)$ ($0\le \bar t\le t$), while in (b)
the system evolves with $H(\bar t)$ for $0\le \bar t\le t$ and $-H(2t-\bar t)$ for $t\le \bar t \le 2t$.}
\label{contour}
\end{center}
\end{figure}

We begin by noting that an out-of-time-ordered function $\langle W(t)V(0)W(t)V(0)\rangle$
is rewritten as
\begin{align}
&
\frac{{\rm Tr}[e^{-\beta H(0)}\, \mathcal U(0,t)\, W\, \mathcal U(t,0)\, V\, \mathcal U(0,t)\, W\, \mathcal U(t,0)\, V]}{{\rm Tr}[e^{-\beta H(0)}]}
\notag
\end{align}
\begin{align}
&=
\frac{{\rm Tr} [\mathcal T_{\tilde{\mathcal C}}\, e^{-i\int_{\tilde{\mathcal C}}dt H(t)} W_{t_+} V_{0_c} W_{t_-} V_{0_-}]}
{{\rm Tr}[\mathcal T_{\tilde{\mathcal C}}\, e^{-i\int_{\tilde{\mathcal C}} dt H(t)}]}
\equiv
\langle \mathcal T_{\tilde{\mathcal C}}\, W_{t_+} V_{0_c}W_{t_-}V_{0_-}\rangle,
\notag
\end{align}
where $\mathcal U(t,t')$
is the unitary evolution operator, $\tilde{\mathcal C}$ is a doubly folded Kadanoff-Baym contour with time running as
$0\, (\equiv 0_-)\to t\,(\equiv t_-)\to 0\,(\equiv 0_c)\to t\,(\equiv t_+)\to 0\,(\equiv 0_+)\to -i\beta$ [see Fig.~\ref{contour}(a)]
\cite{extended_contour},
and $\mathcal T_{\tilde{\mathcal C}}$ is the time ordering operator along $\tilde{\mathcal C}$.
We can also unfold the contour $\tilde{\mathcal C}$ as shown in Fig.~\ref{contour}(b),
where time runs as $0\to 2t\to 0\to -i\beta$. From $t$ to $2t$, the time evolution is reversed, i.e., 
the system evolves with the Hamiltonian $-H(2t-\bar t)$ ($t\le\bar t\le 2t$). These two types of contours are equivalent.
It is straightforward to extend a field theory from $\mathcal C$ to $\tilde{\mathcal C}$. 
The nonequilibrium Green's function is defined on $\tilde{\mathcal C}$ as
$G(t,t')=-i\langle \mathcal T_{\tilde{\mathcal C}}\, c(t)c^\dagger(t')\rangle$ 
[$c$ ($c^\dagger$) is the fermion annihilation (creation) operator].
By replacing $\mathcal C$ with $\tilde{\mathcal C}$,
the nonequilibrium DMFT is generalized as explicitly constructed below.

We consider the FK model with the Hamiltonian
\begin{align}
H(t)&=
\sum_{ij} J_{ij} c_i^\dagger c_j
+\sum_i (-\mu n_i^c+E_f n_i^f)+U(t)\sum_i n_i^f n_i^c,
\end{align}
where $n_i^c\equiv c_i^\dagger c_i$, $n_i^f\equiv f_i^\dagger f_i$,
$J_{ij}$ and $\mu$ are the hopping amplitude and the chemical potential for the $c$ particles, 
$E_f$ is the energy level for the $f$ particles, and $U(t)$ is the on-site interaction, which can be time dependent.
The model is exactly solvable in any dimension in the sense that $[H,n_i^f]=0$
for all $i$, i.e., it has infinitely many conserved quantities. 
The immobile $f$ particles act as a random potential for the itinerant $c$ particles.

In the infinite dimensional limit ($d\to\infty$) with the hopping scaled as
$J_{ij}\propto J_\ast/\sqrt{d}$ ($J_\ast$ is a fixed constant) \cite{MetznerVollhardt1989},
the lattice model can be exactly mapped onto an impurity problem with 
a self-consistently determined dynamical mean field (hybridization function) $\Delta(t,t')$
\cite{BrandtMielsch1989,BrandtMielsch1990,BrandtMielsch1991,GeorgesKotliarKrauthRozenberg1996,FreericksZlatic2003},
where the action is given by
\begin{align}
\mathcal S_{\rm imp}
&=
\int_{\tilde{\mathcal C}} dt dt'\, c^\dagger(t)\Delta(t,t')c(t')
\notag
\\
&\quad
+\int_{\tilde{\mathcal C}} dt\, [-\mu n^c(t)+E_f n^f(t)+U(t)n^f(t)n^c(t)].
\label{S_imp}
\end{align}
The mapping is constructed such that the local lattice Green's function is equal to the impurity Green's function.
In the large-$d$ limit, the lattice self-energy becomes local
and can be identified with the impurity self-energy.
The single-particle Green's function for the FK model can be expressed as
\begin{align}
G(t,t')
&\equiv
-i\langle \mathcal T_{\tilde{\mathcal C}}\, c(t)c^\dagger(t')\rangle_{\mathcal S_{\rm imp}}
=
\sum_\alpha w_\alpha R_\alpha(t,t'),
\label{local G}
\end{align}
where $\alpha=0$ and $1$ correspond to empty and occupied $f$ particle configurations,
$w_1=\langle n^f\rangle$, $w_0=1-w_1$,
and $R_\alpha(t,t')$ is a configuration-dependent Green's function,
which satisfies the Dyson equation,
\begin{align}
[i\partial_t+\mu-U(t)\alpha]R_\alpha(t,t')-\int_{\tilde{\mathcal C}} d\bar t\, \Delta(t,\bar t)R_\alpha(\bar t,t')
&=
\delta_{\tilde{\mathcal C}}(t,t').
\label{Dyson}
\end{align}
Here the integral is taken along the contour $\tilde{\mathcal C}$, and $\delta_{\tilde{\mathcal C}}(t,t')$
is the generalized contour delta function \cite{delta_function}.
$R_0(t,t')$ is the usual Weiss Green's function in the nonequilibrium DMFT \cite{noneqDMFTreview}.
Throughout this Letter, we consider the half-filled case (i.e., $\mu=U/2, w_1=0.5$)
on the Bethe lattice with infinite coordinations, with the density of states
$D(\epsilon)=\sqrt{4J_\ast^2-\epsilon^2}/(2\pi J_\ast^2)$,
and use $J_\ast$ ($J_\ast^{-1}$) as the unit of energy (time). 
In this case, $\Delta(t,t')$ is related to
the local Green's function via $\Delta(t,t')=J_\ast^2 G(t,t')$ \cite{GeorgesKotliarKrauthRozenberg1996}.
Within DMFT, the model shows a metal-to-Mott insulator transition at $U=2$, and the metallic phase is a non-Fermi liquid
\cite{FreericksZlatic2003}.

In the FK model, the impurity action (\ref{S_imp}) can be block-diagonalized 
into $f$-particle configuration sectors ($\alpha=0,1$), in which the $c$ particles behave
as free fermions in an effective potential $U(t)\alpha-\mu$. This makes it possible to calculate
arbitrary local dynamical correlation functions exactly. Let us look at the 2-point charge correlation function,
which is given by the sum of contributions from the two sectors,
\begin{align}
C_2(t_1,t_2)
&\equiv
\langle\mathcal T_{\tilde{\mathcal C}}\, n^c(t_1) n^c(t_2) \rangle
=
\sum_\alpha w_\alpha 
\langle\mathcal T_{\tilde{\mathcal C}}\, n^c(t_1) n^c(t_2) \rangle_{\mathcal S_{\rm imp}^\alpha},
\label{two point1}
\end{align}
where $\mathcal S_{\rm imp}^\alpha
=\int_{\tilde{\mathcal C}} dt dt'\, c^\dagger(t)\Delta(t,t')c(t')
+\int_{\tilde{\mathcal C}} dt\, [U(t)\alpha-\mu] c^\dagger(t)c(t)$
is the sector $\alpha$ ($n^f=\alpha$) of the impurity action.
Since $\mathcal S_{\rm imp}^\alpha$ is quadratic with respect to
the $c$-particle operators, we can analytically evaluate the right-hand side of Eq.~(\ref{two point1})
by Wick's theorem \cite{external_field}. The result is
\begin{align}
C_2(t_1,t_2)
&=
\sum_\alpha w_\alpha 
[R_\alpha(t_1,t_2)R_\alpha(t_2,t_1)-R_\alpha(t_1,t_1)R_\alpha(t_2,t_2)].
\label{two point2}
\end{align}
Here $R_\alpha(t,t')$ at equal times ($t=t'$) should be read as
a lesser component (i.e., $t'=t+0$).
Equation (\ref{two point2}) is symmetric with respect to the exchange $t_1\leftrightarrow t_2$,
and satisfies the boundary condition $\lim_{t_2\to t_1}C_2(t_1,t_2)=\langle n^c(t_1)\rangle$.

As shown in \cite{supplementary}, Eq.~(\ref{two point2}) correctly reproduces
the previous result for the dynamical charge susceptibility
\cite{Shvaika2000,Shvaika2001,FreericksMiller2000,FreericksZlatic2003}
\begin{align}
\chi(t,t')
&=
i\theta(t-t')\langle [n^c(t),n^c(t')]\rangle.
\label{chi}
\end{align}
In the infinite dimensional limit, the local dynamical charge susceptibility
is equal to the lattice charge susceptibility $\chi_{\bm q}(t,t')$ at general momentum $\bm q$
(randomly chosen from the Brillouine zone),
since $\chi(t,t')=N_{\bm q}^{-1}\sum_{\bm q} \chi_{\bm q}(t,t')$ 
and general momenta $\bm q$ make
equal and dominant contributions to the momentum sum.
In fact, we can confirm by explicit calculations that expression (\ref{two point2})
gives the correct dynamical charge susceptibility for general momentum \cite{supplementary}.

The calculation can straightforwardly be generalized to arbitrary $n$-point local charge correlation functions.
The 3-point function is given by
\begin{align}
&
C_3(t_1,t_2,t_3)
\equiv
\langle\mathcal T_{\tilde{\mathcal C}}\, n^c(t_1) n^c(t_2) n^c(t_3)\rangle
\notag
\\
&=
\sum_\alpha w_\alpha
\{
i[R_\alpha(t_1,t_2)R_\alpha(t_2,t_3)R_\alpha(t_3,t_1)
+(\mbox{1 term})]
\notag
\\
&\quad
-i[R_\alpha(t_1,t_1)R_\alpha(t_2,t_3)R_\alpha(t_3,t_2)
+(\mbox{2 terms})]
\notag
\\
&\quad
+iR_\alpha(t_1,t_1)R_\alpha(t_2,t_2)R_\alpha(t_3,t_3)
\},
\label{three point}
\end{align}
and the 4-point function by
\begin{align}
&
C_4(t_1,t_2,t_3,t_4)
\equiv
\langle\mathcal T_{\tilde{\mathcal C}}\, n^c(t_1) n^c(t_2) n^c(t_3) n^c(t_4)\rangle
\notag
\\
&=
\sum_\alpha
w_\alpha\{
-[R_\alpha(t_1,t_2)R_\alpha(t_2,t_3)R_\alpha(t_3,t_4)R_\alpha(t_4,t_1)
+(\mbox{5 terms})]
\notag
\\
&\quad
+[R_\alpha(t_1,t_1)R_\alpha(t_2,t_3)R_\alpha(t_3,t_4)R_\alpha(t_4,t_2)
+(\mbox{7 terms})]
\notag
\\
&\quad
-[R_\alpha(t_1,t_1)R_\alpha(t_2,t_2)R_\alpha(t_3,t_4)R_\alpha(t_4,t_3)
+(\mbox{5 terms})]
\notag
\\
&\quad
+[R_\alpha(t_1,t_2)R_\alpha(t_2,t_1)R_\alpha(t_3,t_4)R_\alpha(t_4,t_3)
+(\mbox{2 terms})]
\notag
\\
&\quad
+R_\alpha(t_1,t_1)R_\alpha(t_2,t_2)R_\alpha(t_3,t_3)R_\alpha(t_4,t_4)
\}.
\label{four point}
\end{align}
In Eqs.~(\ref{three point}) and (\ref{four point}), we group the terms by topologically equivalent Wick contractions,
and in each group one representative term is spelt out.
As a consistency check, one can see that the right-hand sides of Eqs.~(\ref{three point}) and (\ref{four point})
are invariant under arbitrary permutations $t_i \leftrightarrow t_j$. We can also confirm that 
these correlation functions satisfy the boundary conditions,
$\lim_{t_3\to t_2}C_3(t_1,t_2,t_3)=C_2(t_1,t_2)$
and $\lim_{t_4\to t_3}C_4(t_1,t_2,t_3,t_4)=C_3(t_1,t_2,t_3)$.

The OTOC function (\ref{OTOC}) for $W=V=n^c$ can be expressed as
\begin{align}
C(t)
&=
-C_4(t_+,0_c,t_-,0_-)
-C_4(0_+,t_+,0_c,t_-)
\notag
\\
&\quad
+C_3(t_+,0_c,t_-)+C_3(0_+,t_+,0_c),
\label{OTOC 4 point}
\end{align}
where $t_\pm, 0_\pm, 0_c$ are the time points on $\tilde{\mathcal C}$ defined in Fig.~\ref{contour}
(we can also calculate the OTOC function for $W=c$, $V=c^\dagger$ \cite{supplementary}).
Let us emphasize that the result (\ref{OTOC 4 point}) is valid not only in equilibrium but also out of equilibrium.
For details of the numerical implementation, we refer to Ref.~\cite{noneqDMFTreview}.

\begin{figure}[t]
\begin{center}
\includegraphics[width=7cm]{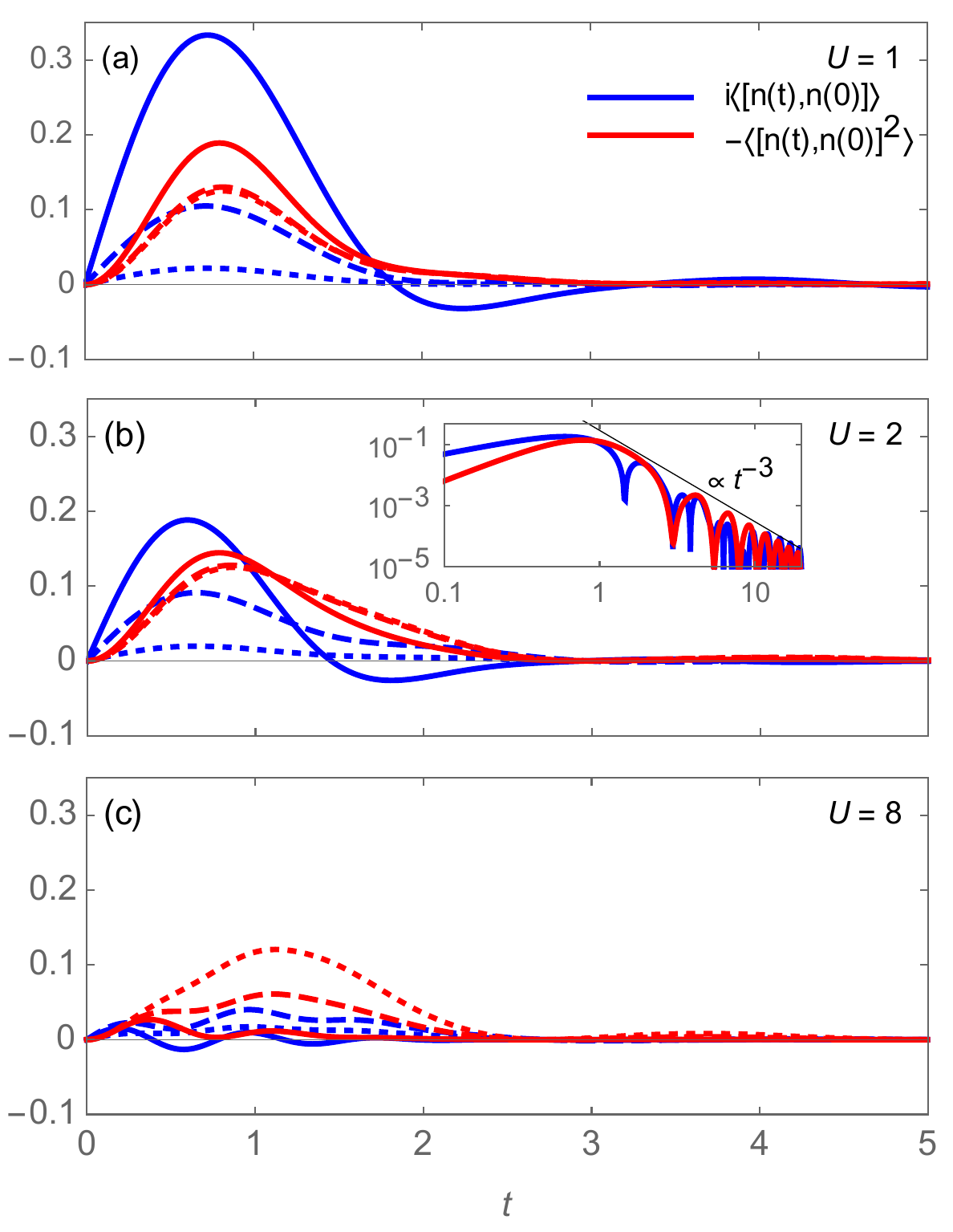}
\caption{Dynamical charge susceptibility $\chi(t)$ (blue curves) and out-of-time-ordered charge correlation function
$C(t)$ (red) for the FK model
with $U=1$ (a), $U=2$ (b), and $U=8$ (c). The solid, dashed, and dotted curves correspond to 
$\beta=10$, $0.5$, and $0.1$, respectively.
The inset shows the corresponding log-log plot for the absolute values, 
compared with the asymptotic behavior $\propto t^{-3}$.}
\label{Fig:otoc}
\end{center}
\end{figure}

The results for the OTOC function $C(t)=-\langle [n^c(t),n^c(0)]^2\rangle$ are shown in Fig.~\ref{Fig:otoc} (red curves)
for several $U$ and $\beta$.
As a comparison, we also plot the dynamical charge susceptibility $\chi(t)\equiv\chi(t,0)$ [Eq.~(\ref{chi})] (blue), which is a usual response function obeying causality.
Both $C(t)$ and $\chi(t)$ grow after $t=0$, peak out within $t\lesssim 1$, and gradually decay to zero. By definition, $C(t)\ge 0$, while $\chi(t)$ oscillates around zero.
Initially the correlations build up as $C(t)\propto t^2$ and $\chi(t)\propto t$. Both of them show 
a long-time asymptotic behavior $\sim t^{-3}$, reflecting
the power-law decay of the Green's function, $R_\alpha^R(t,0)\sim t^{-3/2}$ \cite{supplementary}.

In the SYK model or other systems that show the AdS/CFT correspondence, 
one finds a separation of the relevant time scales; 
the time scale of the change of $C(t)$ (scrambling time)
is longer than that of
ordinary response functions such as $\chi(t)$ (thermalization time)
by a factor of $\log N$ \cite{MaldacenaShenkerStanford2016},
where  $N$ is the number of sites for the SYK model or the number of colors in CFTs.
In contrast, there is no clear separation of the two time scales in the FK model (Fig.~\ref{Fig:otoc}),
indicating that the AdS/CFT correspondence cannot be applied here.
Given this circumstance, we do not clearly see an exponential growth (butterfly effect) 
of the deviation of the OTOC from the initial value.
In this sense, the FK model does not describe a chaotic system, which is consistent with the expectation that
systems with infinitely many conserved charges are not chaotic (and do not thermalize \cite{EcksteinKollar2008}).

\begin{figure}[t]
\begin{center}
\includegraphics[width=7cm]{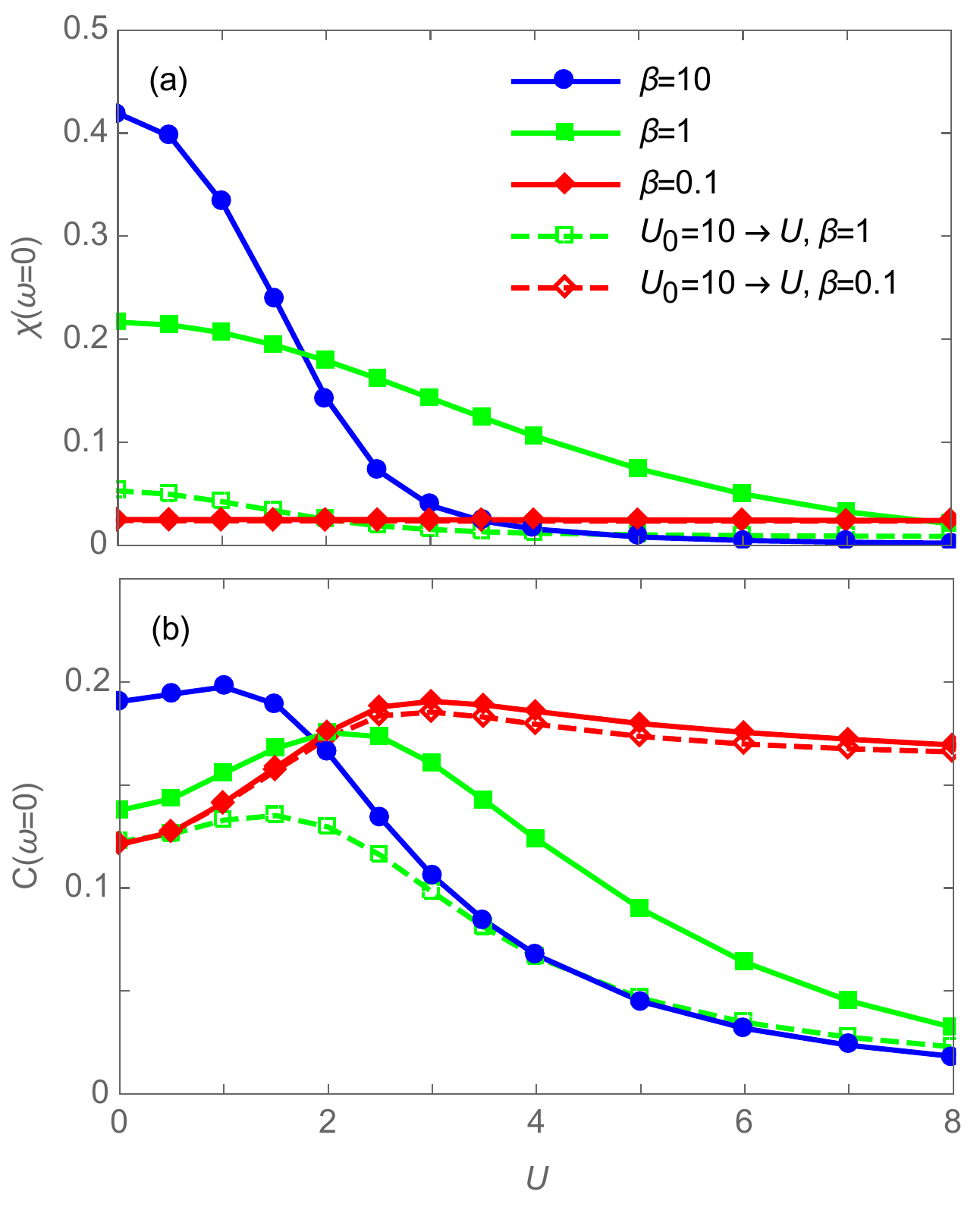}
\caption{Comparison between the dynamical charge susceptibility $\chi(\omega)$ (a) and 
the OTOC spectral function 
$C(\omega)$ (b) at $\omega=0$ for the FK model. The dashed curves correspond to interaction quenches $U_0=10\to U$
with initial temperatures $\beta=1$ and $0.1$.}
\label{Fig:otoc-spectrum}
\end{center}
\end{figure}

Nevertheless, if we look at the temperature dependence of $C(t)$ in Fig.~\ref{Fig:otoc},
it is very different from that of the ordinary response function $\chi(t)$. As the temperature increases,
the amplitude of $\chi(t)$ vanishes irrespective of $U$,
whereas $C(t)$ is enhanced in the insulating phase [$U>2$, Fig.~\ref{Fig:otoc}(c)], 
more or less unchanged at the critical point [$U=2$, Fig.~\ref{Fig:otoc}(b)],
and suppressed in the metallic phase [$U<2$, Fig.~\ref{Fig:otoc}(a)].
In particular, $C(t)$ remains nonvanishing and interaction-dependent in the high-temperature limit,
even though the dynamical charge correlations are completely suppressed ($\chi(t)\to 0$). 

To quantify the temperature and interaction dependence of the OTOC at low energy, we compute
the OTOC spectral function
$C(\omega)=\int_0^\infty dt\, e^{i\omega t}C(t)$
and the dynamical charge susceptibility spectrum
$\chi(\omega)=\int_0^\infty dt\, e^{i\omega t}\chi(t,0)$.
$C(\omega=0)$ measures how much the OTOC grows during the entire dynamics.
Figure~\ref{Fig:otoc-spectrum} plots $\chi(\omega=0)$ and $C(\omega=0)$ for several $U$ and $\beta$.
$\chi(\omega=0)$ monotonically decreases as a function of $U$ at arbitrary fixed temperature.
At sufficiently low temperature, the charge gap opens in the insulating phase ($U>2$), where $\chi(\omega=0)$ vanishes.
As the temperature increases, thermal excitations take place above the charge gap,
leading to an increase in $\chi(\omega=0)$
in the insulating phase. If the temperature further increases, charge correlations disappear
and $\chi(\omega)$ approaches zero. In contrast, $C(\omega=0)$ is a non-monotonic function of $U$;
It reaches the maximum
at some intermediate coupling ($1\lesssim U\lesssim 3$) around the metal-insulator transition point
[Fig.~\ref{Fig:otoc-spectrum}(b)], 
and decays to zero in the $U\to\infty$ limit.
In particular, the high-temperature limit of $C(\omega=0)$,
which is close to that of $\beta=0.1$ in Fig.~\ref{Fig:otoc-spectrum}(b), shows a highly non-trivial 
non-zero spectral weight amplified in the insulating phase,
whereas the ordinary response function becomes trivial. The overall temperature dependence of $C(\omega=0)$
is similar to that of $C(t)$ discussed in the preceding paragraph.

\begin{figure}[t]
\begin{center}
\includegraphics[width=7cm]{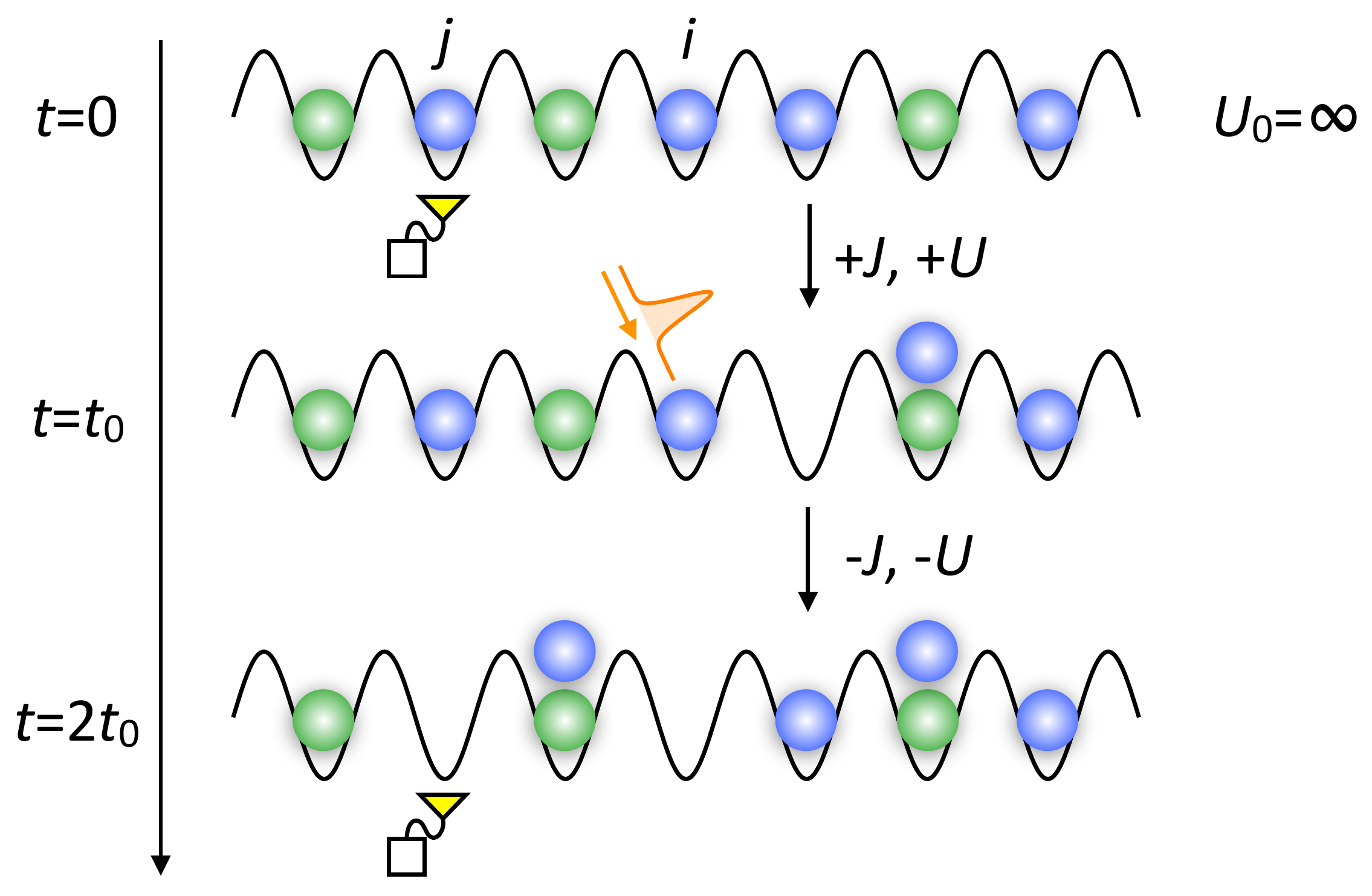}
\caption{Illustration of the proposed measurement protocol of the OTOC in ultracold atomic systems
with two fermionic species (blue and green). At $t=t_0$, a local pulse is applied to site $i$, and at $t=0$ and $t=2t_0$
the particle density is measured at site $j$.}
\label{experiment}
\end{center}
\end{figure}

Finally, let us discuss how to experimentally measure the OTOC $C(t)$ in many fermion systems.
We consider ultracold atomic systems in an optical lattice. 
There have been several proposals on the measurement of OTOCs.
One strategy is to take an interferometric approach with a qubit control \cite{Swingle2016,Danshita2016,Yao2016,Zhu2016}.
Another approach is a time-reversal protocol \cite{Garttner2016,Li2016}.
Since ultracold atomic systems offer full control over the Hamiltonian parameters with negligible dissipation on the time scale of interest,
we propose, based on the latter approach, 
a serial protocol (Fig.~\ref{experiment}) implemented along the contour in Fig.~\ref{contour}(b)
that is feasible with available experimental techniques for atomic gases.

We prepare a Mott insulating state with the initial $U_0=\infty$ tuned
by a Feshbach resonance, and measure the number ($\equiv N_1$) of $c$ particles at site $j$ nondestructively.
(Note that the Mott insulator is an eigenstate of the particle density.) Then we quench the interaction $U_0=\infty\to U$
at $t=0$, and let the system time-evolve for a duration of $t_0$.
At $t=t_0$, we apply an instantaneous local potential pulse $\delta H(t)=\varepsilon\delta(t-t_0)n_i^c(t)$
by using a focused laser to shift the energy level
of site $i$ \cite{pulse}. 
After that, we change the hopping $J_{ij}\to -J_{ij}$ by shaking the lattice periodically \cite{TsujiOkaAoki2009}, or applying a $\pi$ pulse \cite{TsujiOkaAokiWerner2012}, or using a Raman process \cite{JakschZoller2005}
to induce a $\pi$ phase shift in the kinetic term \cite{heating}. 
At the same time, we quench the interaction $U\to -U$. With these, we can flip the sign of the Hamiltonian, 
$H\to -H$ \cite{periodic},
which enables the system to effectively propagate backward in time 
(similar to a Loschmidt-echo experiment).
After letting the system evolve for another $t_0$, we measure the $c$ particle density ($\equiv N_2$) at site $j$.
We repeat this procedure to measure the expectation value of $2N_1N_2-N_2$ \cite{parity}. 
If we expand it with respect to $\varepsilon$, we obtain
\begin{align}
&
\langle 2N_1N_2-N_2\rangle
\notag
\\
&=
{\rm Tr}[\rho(0) \mathcal U(-t_0)e^{i\varepsilon n_i^c}\mathcal U(t_0)n_j^c
\mathcal U(-t_0)e^{-i\varepsilon n_i^c}\mathcal U(t_0)(2n_j^c-1)]
\notag
\\
&=
\langle n_j^c(0)\rangle
+\varepsilon^2
\langle [n_i^c(t_0),n_j^c(0)]^2\rangle
+O(\varepsilon^3),
\end{align}
where $\rho(0)$ is the initial density matrix, and $\mathcal U(t,t')\equiv\mathcal U(t-t')$.
The leading-order contribution is exactly the OTOC $C(t)$ for the nonequilibrium system
subject to the interaction quench $U_0\to U$.
In Fig.~\ref{Fig:otoc-spectrum} (dashed curves), we show $\chi(\omega=0)$ and $C(\omega=0)$ for the interaction quench 
where the initial interaction is chosen to be large but finite ($U_0=10$). We can see
the non-monotonic behavior and characteristic temperature dependence of $C(\omega=0)$,
similar to the results obtained in equilibrium.
This scheme is applicable not only to the FK model but also to the Hubbard model.
To realize the FK model in ultracold atomic systems, we need a mass imbalance between
two fermionic species.

To summarize, we have obtained an exact solution for the OTOCs of the FK model in the thermodynamic limit
by generalizing the nonequilibrium DMFT to the extended Kadanoff-Baym contour $\tilde{\mathcal C}$.
We find that the OTOC $C(t)$ is most enhanced around the metal-insulator transition point,
and remains nontrivial in the high-temperature limit,
which can be measured in ultracold atomic systems.
Our work is a first step toward an analysis of OTOCs in more complicated fermionic lattice models such as the Hubbard model.
It is of interest to investigate whether those models, especially in the ``strange-metal'' phase,
show a chaotic behavior, and if so how fast they are scrambled.

We thank I. Danshita, T. Fukuhara, and I. Nakamura for stimulating discussions.
NT is supported by JSPS KAKENHI Grant No. JP16K17729.
PW acknowledges support from ERC starting grant No. 278023.
MU acknowledges support by KAKENHI Grant No. JP26287088 and KAKENHI Grant No. JP15H05855.

\bibliographystyle{apsrev}
\bibliography{ref}

\begin{thebibliography}{61}
\bibitem[{del()}]{delta_function}
\bibinfo{note}{The generalized contour delta function on $\tilde{\mathcal C}$
  is defined by $\int_{\tilde{\mathcal C}} d\bar t\, \delta_{\tilde{\mathcal
  C}}(t,\bar t)g(\bar t)=g(t)$ for an arbitrary contour function $g(t)$.}
\bibitem[{ext({\natexlab{b}})}]{external_field}
\bibinfo{note}{One can also derive the same result by introducing an external
  field that couples to the $c$ particle density and taking a functional
  derivative of the density with respect to the field.}
\bibitem[{pul()}]{pulse}
\bibinfo{note}{In reality, the local pulse will not be perfectly delta-function
  like. If one uses a pulse shape $f(t,\bm r)$ with finite widths in time and
  space, we obtain the sum of contributions $\int dt'\int d\bm r' f(t-t',\bm
  r_i-\bm r')\langle [n_{\bm r'}^c(t'),n_j^c(0)]^2\rangle$.}
\bibitem[{hea()}]{heating}
\bibinfo{note}{During the sign reversal of the hopping, there might occur
  heating. In this respect, the $\pi$ pulse method is superior to others since
  the system is driven only in a short time interval.}
\bibitem[{per()}]{periodic}
\bibinfo{note}{If one uses a periodic drive, one has to quench $U\to
  -|J_0(A)|U$ [$J_0(A)$ is the zeroth-order Bessel function and $A$ is the
  amplitude of the drive] so that the ratio between the hopping and interaction
  remains unchanged. In this case, the time after $t=t_0$ should be rescaled
  accordingly.}
\bibitem[{par()}]{parity}
\bibinfo{note}{Since $N_1,N_2=0,1$ for fermionic systems, it is enough to
  measure the parity of the particle density.}
\end{thebibliography}

\end{document}


\title{
Supplementary Material for
``Exact Out-of-Time-Ordered Correlation Functions for an Interacting Lattice Fermion Model''
}

\author{Naoto Tsuji}
\affiliation{RIKEN Center for Emergent Matter Science (CEMS), Wako 351-0198, Japan}
\author{Philipp Werner}
\affiliation{Department of Physics, University of Fribourg, 1700 Fribourg, Switzerland}
\author{Masahito Ueda}
\affiliation{RIKEN Center for Emergent Matter Science (CEMS), Wako 351-0198, Japan}
\affiliation{Department of Physics, University of Tokyo, Hongo, Tokyo 113-0033, Japan}





\maketitle

\section{I. \hspace{.1cm} Equivalence of the local and lattice susceptibilities at general momentum}

In this section, we explicitly show that the local dynamical charge susceptibility (8)
of the Falicov-Kimball (FK) model,
calculated from Eq.~(7), is equal to the previously known result
for the lattice dynamical charge susceptibility at general momentum 
\cite{FreericksMiller2000,FreericksZlatic2003}. 

To see this, we substitute Eq.~(7) in Eq.~(8) to obtain
\begin{align}
\chi(t,t')
&=
i\theta(t-t')\sum_\alpha w_\alpha [R_\alpha^>(t,t')R_\alpha^<(t',t)-R_\alpha^<(t,t')R_\alpha^>(t',t)].
\end{align}
We replace the greater Green's functions with the lesser, retarded, and advanced components by
using the relations
\begin{align}
\theta(t-t')R_\alpha^>(t,t')
&=
\theta(t-t')R_\alpha^<(t,t')+R_\alpha^R(t,t'),
\\
\theta(t-t')R_\alpha^>(t',t)
&=
\theta(t-t')R_\alpha^<(t',t)-R_\alpha^A(t',t).
\end{align}
This results in the expression
\begin{align}
\chi(t,t')
&=
i\sum_\alpha w_\alpha [R_\alpha^R(t,t')R_\alpha^<(t',t)+R_\alpha^<(t,t')R_\alpha^A(t',t)],
\end{align}
or, in the Fourier transformed form,
\begin{align}
\chi(\omega)
&=
i\int \frac{d\omega'}{2\pi}\sum_\alpha w_\alpha 
[R_\alpha^R(\omega+\omega')R_\alpha^<(\omega')+R_\alpha^<(\omega+\omega')R_\alpha^A(\omega')].
\label{chi omega}
\end{align}
By means of the fluctuation-dissipation relation 
$R_\alpha^<(\omega)=f(\omega)[R_\alpha^A(\omega)-R_\alpha^R(\omega)]$ 
[$f(\omega)=1/(e^{\beta\omega}+1)$ is the Fermi distribution function], 
Eq.~(\ref{chi omega}) can be written as
\begin{align}
\chi(\omega)
&=
-i\int \frac{d\omega'}{2\pi}\sum_\alpha w_\alpha \bigg\{
f(\omega')R_\alpha^R(\omega+\omega')R_\alpha^R(\omega')
\notag
\\
&\quad
-f(\omega+\omega')R_\alpha^A(\omega+\omega')R_\alpha^A(\omega')
\notag
\\
&\quad
-[f(\omega')-f(\omega+\omega')]R_\alpha^R(\omega+\omega')R_\alpha^A(\omega')
\bigg\}.
\label{chi omega2}
\end{align}

Let us recall that the configuration-dependent Green's function $R_\alpha(t,t')$ is in a nontrivial way related to
the local Green's function and the self-energy \cite{FreericksMiller2000},
\begin{align}
\frac{1}{G^X(\omega)G^Y(\omega')}-\frac{1}{\sum_\alpha w_\alpha R_\alpha^X(\omega)R_\alpha^Y(\omega')}
&=
\frac{\Sigma^X(\omega)-\Sigma^Y(\omega')}{G^X(\omega)-G^Y(\omega')},
\label{FM relation}
\end{align}
where $X,Y=R,A$. The right-hand side of Eq.~(\ref{FM relation}) is the irreducible dynamical charge vertex function.
The relation (\ref{FM relation}) can be proven as follows:
\begin{align}
&
\sum_\alpha w_\alpha R_\alpha^X(\omega)R_\alpha^Y(\omega')
=
\sum_\alpha w_\alpha \frac{R_\alpha^X(\omega)-R_\alpha^Y(\omega')}{R_\alpha^Y{}^{-1}(\omega')-R_\alpha^X{}^{-1}(\omega)}
\notag
\\
&=
\frac{\sum_\alpha w_\alpha [R_\alpha^X(\omega)-R_\alpha^Y(\omega')]}{R_0^Y{}^{-1}(\omega')-R_0^X{}^{-1}(\omega)}
\notag
\\
&=
\frac{G^X(\omega)-G^Y(\omega')}{[G^Y{}^{-1}(\omega')+\Sigma^Y(\omega')]-[G^X{}^{-1}(\omega)+\Sigma^X(\omega)]}.
\label{wRR}
\end{align}
In deriving the second equality, we used $R_1^X{}^{-1}(\omega)=R_0^X{}^{-1}(\omega)-U$.
In deriving the last equality, we used the impurity solution (4) and
the impurity Dyson equation $G^X(\omega)=[R_0^X{}^{-1}(\omega)-\Sigma^X(\omega)]^{-1}$
[note that $R_0^X(\omega)$ is the Weiss Green's function].
One immediately gets Eq.~(\ref{FM relation}) from the last line of Eq.~(\ref{wRR}).
Substituting Eq.~(\ref{FM relation}) into Eq.~(\ref{chi omega2}), we arrive at
\begin{align}
&\chi(\omega)
=
-i\int \frac{d\omega'}{2\pi}
\bigg\{ f(\omega')
\frac{G^R(\omega+\omega')G^R(\omega')}{1
-G^R(\omega+\omega')G^R(\omega')
\frac{\Sigma^R(\omega+\omega')-\Sigma^R(\omega')}{G^R(\omega+\omega')-G^R(\omega')}}
\notag
\\
&
-f(\omega+\omega')
\frac{G^A(\omega+\omega')G^A(\omega')}{1
-G^A(\omega+\omega')G^A(\omega')
\frac{\Sigma^A(\omega+\omega')-\Sigma^A(\omega')}{G^A(\omega+\omega')-G^A(\omega')}}
\notag
\\
&
-[f(\omega')-f(\omega+\omega')]
\frac{G^R(\omega+\omega')G^A(\omega')}{1
-G^R(\omega+\omega')G^A(\omega')
\frac{\Sigma^R(\omega+\omega')-\Sigma^A(\omega')}{G^R(\omega+\omega')-G^A(\omega')}}
\bigg\}.
\end{align}
This is nothing but the lattice dynamical charge susceptibility of the FK model
at general momentum \cite{FreericksMiller2000,FreericksZlatic2003}
[for the hypercubic lattice the general momentum corresponds to
$X(\bm q)=\lim_{d\to\infty}\sum_{i=1}^d\cos q_i/d=0$ \cite{GeorgesKotliarKrauthRozenberg1996,FreericksZlatic2003}]. 
The difference of the sign is due to the different
definitions of the charge susceptibility. The physical reason for the coincidence between
the local and lattice susceptibilities is explained in the main text.

\section{II. \hspace{.1cm} Long-time asymptotic behavior of the out-of-time-ordered correlation function}

In this section, we explain the long-time asymptotic form of the out-of-time-ordered correlation function
$C(t)=-\langle [n^c(t),n^c(0)]^2\rangle$ for the FK model. 

Here $C(t)$ is determined from Eqs.~(9), (10), and (11) in the main text.
We note that $C(t)$ is obtained as a sum of products of $R_\alpha(t,t')$.
Since $R_\alpha(t,t')$ is a 2-point function, any $R_\alpha(t,t')$ with $t,t'\in\tilde{\mathcal C}$
reduces to $R_\alpha(t,t')$ with $t,t'\in\mathcal C$.
For example, $R_\alpha(t_+,0_c)=R_\alpha^>(t,0)$ and $R_\alpha(t_-,0_c)=R_\alpha^<(t,0)$.
Thus $C(t)$ can be expressed as a contour function defined on the conventional Kadanoff-Baym contour $\mathcal C$.
The result is
\begin{align}
C(t)
&=
\sum_\alpha w_\alpha |R_\alpha^R(t,0)|^2
\Big[|R_\alpha^>(t,0)|^2+|R_\alpha^<(t,0)|^2-|R_\alpha^R(t,0)|^2
\notag
\\
&\quad
+2R_\alpha^<(t,t)R_\alpha^<(0,0)-iR_\alpha^<(t,t)-iR_\alpha^<(0,0)\Big].
\label{C(t)}
\end{align}

\begin{figure}[t]
\begin{center}
\includegraphics[width=7cm]{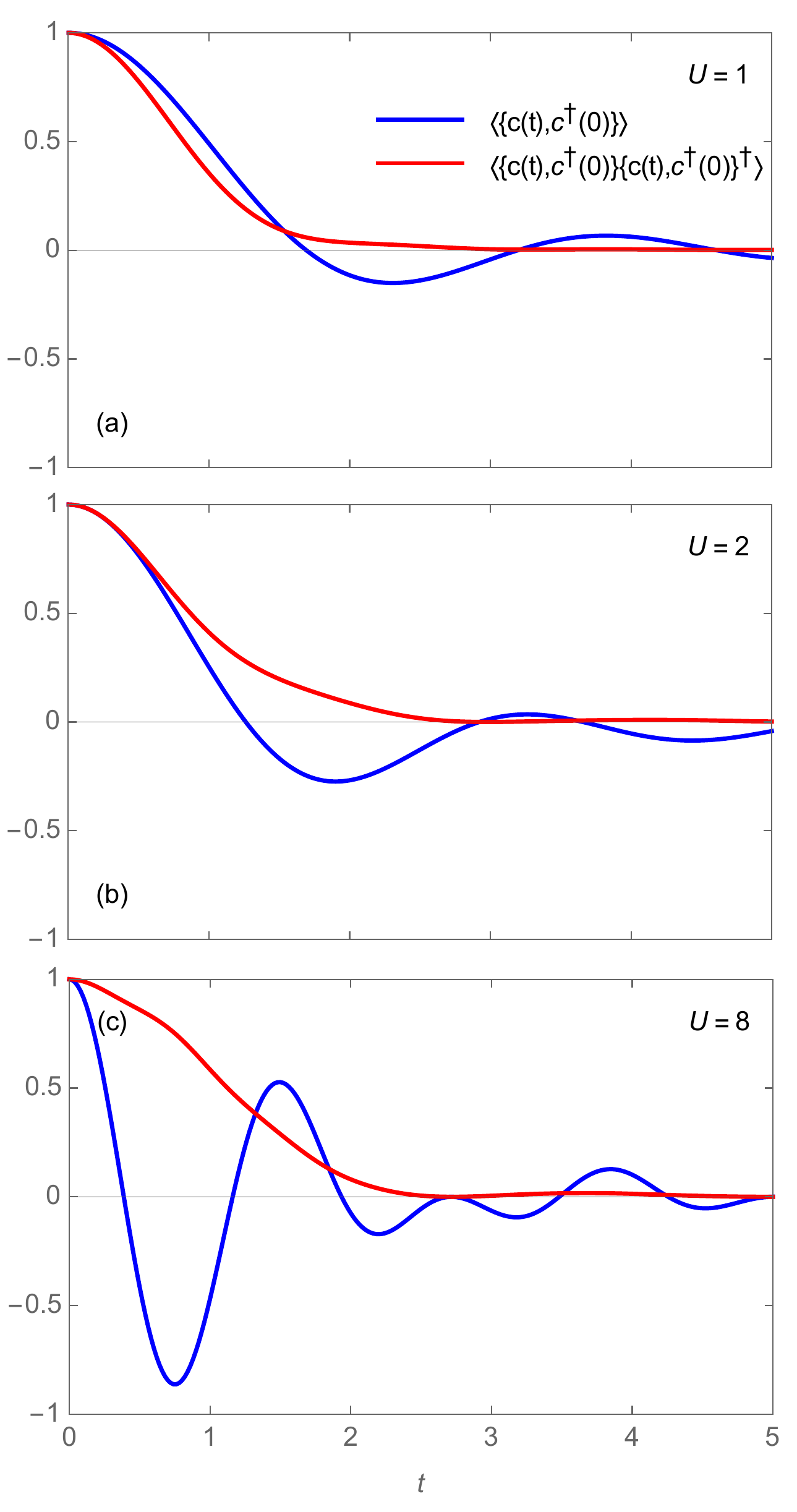}
\caption{Comparison between the retarded Green's function ($\times i$) $\langle \{c(t),c^\dagger(0)\}\rangle$ (blue curves) and 
the out-of-time-ordered correlation function $\langle \{c(t),c^\dagger(0)\}\{c(t),c^\dagger(0)\}^\dagger\rangle$ (red)
for the FK model with $U=1$ (a), $U=2$ (b), and $U=8$ (c). These functions
do not depend on temperature.}
\label{otoc-c}
\end{center}
\end{figure}

The long-time behavior of $|R_\alpha^{R,\gtrless}(t,0)|$ is determined by the branch points
in $R_\alpha^{R,\gtrless}(\omega)$ at the spectral edges ($\omega=\omega_c$), i.e., 
$R_\alpha^{R,\gtrless}(\omega)\sim (\mbox{regular part})+(\mbox{const.})\times \!\!\sqrt{\omega^2-\omega_c^2}$
($\omega\sim\omega_c$). By the saddle point approximation, one finds that 
$|R_\alpha^{R,\gtrless}(t,0)|\sim t^{-3/2}$ in the long-time limit both in the metallic and insulating phases
(although the number of branch points changes at the phase transition point).
This is also confirmed numerically. 
[If one takes a lattice other than the Bethe lattice, the asymptotic behavior may change
according to the form of $R_\alpha^{R,\gtrless}(\omega)$ at the spectral edges.]

Since $R_\alpha^<(t,t)$ approaches a constant for $t\to\infty$, the long-time behavior
of $C(t)$ is governed by $|R_\alpha^R(t,0)|^2 \sim t^{-3}$.
The terms in the fourth power of $R_\alpha$ in Eq.~(\ref{C(t)}) give subleading contributions
to the long-time behavior.

\section{III.\hspace{.1cm} Fermionic out-of-time-ordered correlation function}

In this section, we show the results for the out-of-time-ordered correlation function
constructed from the fermionic operators $V=c_i^\dagger, W=c_i$. A natural extension of the definition
of the OTOC (1) to fermionic operators is given by $\langle \{W(t),V(0)\}\{W(t),V(0)\}^\dagger\rangle$,
where we have replaced the commutator in Eq.~(1) with an anticommutator.
This fermionic OTOC can be evaluated exactly for the infinite-dimensional
FK model with the same technique as explained in the main text:
\begin{align}
C_f(t)
&\equiv
\langle \{c(t),c^\dagger(0)\}\langle \{c(t),c^\dagger(0)\}^\dagger\rangle
\notag
\\
&=
\sum_\alpha w_\alpha [R_\alpha^>(t,0)-R_\alpha^<(t,0)][R_\alpha^<(0,t)-R_\alpha^>(0,t)]
\notag
\\
&=
\sum_\alpha w_\alpha |R_\alpha^R(t,0)|^2.
\end{align}
It satisfies $C_f(t)\ge 0$.

\begin{figure}[t]
\begin{center}
\includegraphics[width=7cm]{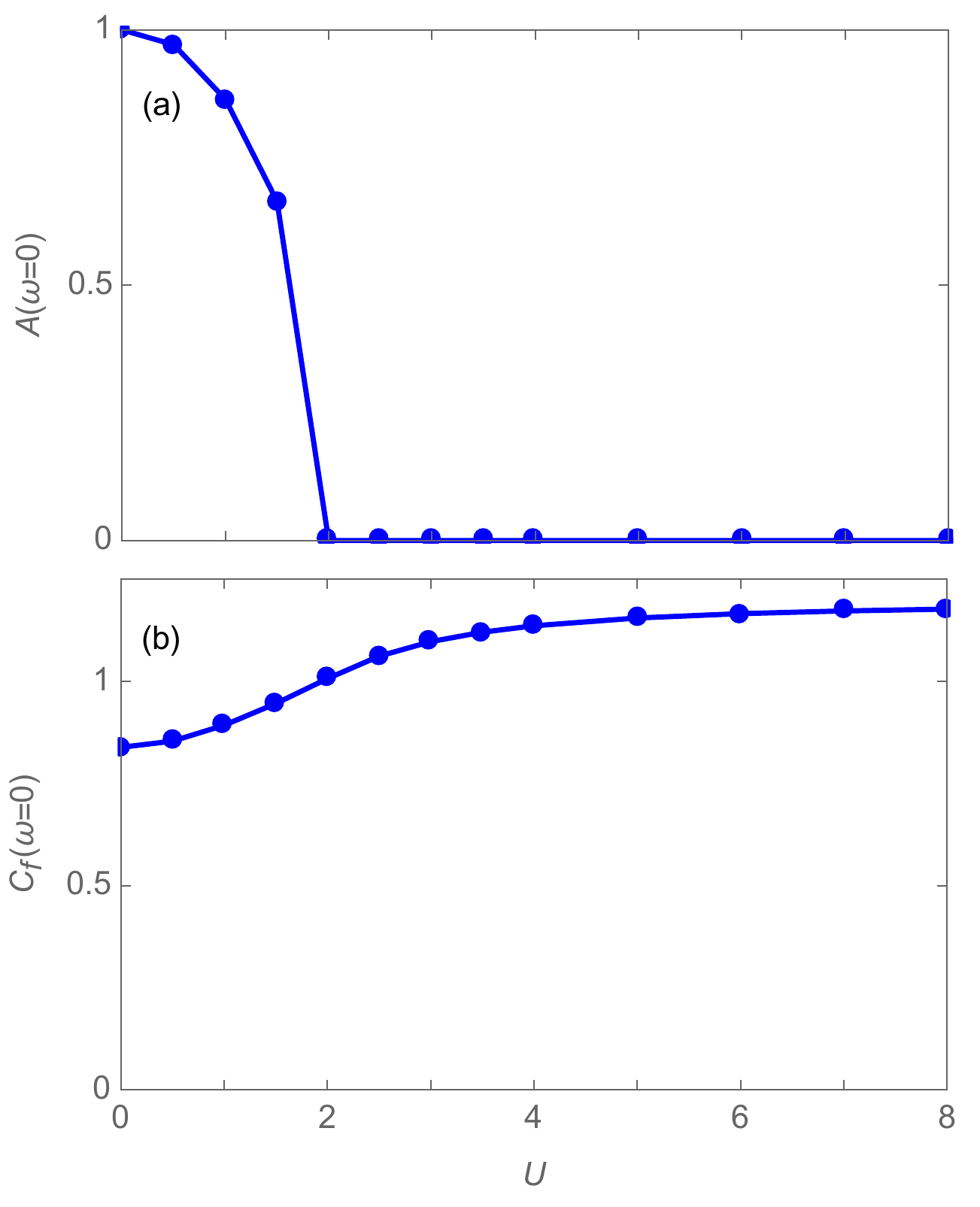}
\caption{Single-particle [$A(\omega)$ (a)] and OTOC [$C_f(\omega)$ (b)] spectral functions for the FK model at zero frequency.
These functions do not depend on temperature.}
\label{otoc-c-spectrum}
\end{center}
\end{figure}

In Fig.~\ref{otoc-c}, we compare the retarded Green's function $iG^R(t,0)$ and $C_f(t)$ for several $U$.
We note that both $G^R(t,0)$ and $C_f(t)$ are independent of the system's temperature, since
the retarded components of Green's functions and the self-energy form a closed set of 
self-consistent equations in the case of the FK model, and can be determined independently of the Matsubara components.
As one can see in Fig.~\ref{otoc-c}, the timescale on which $C_f(t)$ changes is comparable to that of $G^R(t,0)$
(and the dynamical charge correlation function). The initial drop of $C_f(t)$ is delayed compared with $G^R(t,0)$
in the insulating phase ($U\ge 2$). In the long-time limit, the functions decay as power laws:
$iG^R(t,0)\sim t^{-3/2}$ and $C_f(t)\sim t^{-3}$.

To illustrate the dependence of the OTOC $C_f(t)$ on the interaction $U$, we define the OTOC spectral function
\begin{align}
C_f(\omega)=\int_0^\infty dt\, e^{i\omega t} C_f(t).
\end{align}
In Fig.~\ref{otoc-c-spectrum}, we compare $C_f(\omega)$ at $\omega=0$ with the single-particle spectral function
$A(\omega)=-\frac{1}{\pi}{\rm Im}G^R(\omega)$ at $\omega=0$ for the FK model. In the metallic phase ($U<2$), 
there is a nonzero spectral weight at the Fermi energy, while in the insulating phase ($U>2$)
the energy gap opens and the spectral weight vanishes. On the other hand,
$C_f(\omega=0)$ grows as $U$ is increased, and saturates in the strong-coupling limit.
This implies that in the insulating phase, even though the single-particle motions are frozen at low energy,
the information spreading still occurs to some extent.

\bibliographystyle{apsrev}
\bibliography{ref}